\documentclass[twocolumn,showpacs,preprintnumbers,amsmath,amssymb,prl]{revtex4}

\usepackage{graphicx}
\usepackage{dcolumn}
\usepackage{bm}
\usepackage[]{epsfig}

\newcommand{\taur}{\tau_{\mathrm {r}}}

\begin{document}
\title{Resolving long-range spatial correlations in jammed colloidal
    systems using photon correlation imaging}

\author{A. Duri$^{1,2}$, D.A. Sessoms$^3$, V. Trappe$^3$, and L.
Cipelletti$^{1,*}$}

\affiliation{$^1$LCVN, UMR 5587 Universit\'{e}
Montpellier 2 and CNRS, France \\
$^2$IATE, UMR 1208, INRA-CIRAD-UMII-Supagro, France\\
$^3$D\'{e}partement de Physique Universit\'{e} de Fribourg,
Suisse\\}

\email{lucacip@lcvn.univ-montp2.fr}
\date{\today}

\begin{abstract}
We introduce a new dynamic light scattering method, termed photon
correlation imaging, which enables us to resolve the dynamics of
soft matter in space and time. We demonstrate photon correlation
imaging by investigating the slow dynamics of a quasi
two-dimensional coarsening foam made of highly packed, deformable
bubbles and a rigid gel network formed by dilute, attractive
colloidal particles. We find the dynamics of both systems to be
determined by intermittent rearrangement events. For the foam, the
rearrangements extend over a few bubbles, but a small dynamical
correlation is observed up to macroscopic length scales. For the
gel, dynamical correlations extend up to the system size. These
results indicate that dynamical correlations can be extremely
long-ranged in jammed systems and point to the key role of
mechanical properties in determining their nature.

\end{abstract}

\pacs{64.70.pv,82.70.-y,82.70.Rr}
\maketitle

The slow relaxation of molecular fluids, colloidal suspensions and
granular materials becomes increasingly
heterogeneous~\cite{EdigerReview,GlotzerReview,RichertReview,WeeksScience2000,DauchotPRL2005_2}
as the glass~\cite{Donth2001} or jamming~\cite{LiuNature1998}
transitions is approached. In these systems, structural relaxation
occurs through the correlated motion of clusters of
particles, leading to spatial correlations of the local dynamics.
Simulations~\cite{GlotzerReview} and
experiments~\cite{EdigerReview,RichertReview,WeeksScience2000,BerthierScience2005,PREAlba}
of molecular and colloidal supercooled fluids show that the range of
dynamical correlations is typically moderate, extending over at most a few particles
close to the glass transition. Experiments in driven granular
systems~\cite{DauchotPRL2005_2,Durian} support similar conclusions
for athermal systems approaching jamming. However, it has been
recently argued~\cite{PicardPRE2005} that in jammed materials any
local rearrangement should propagate elastically over larger
distances, leading to long-ranged dynamical correlations.

Testing these ideas is a challenging task. Spatial correlations of
the dynamics can of course be investigated in simulations, but the
linear size of the simulation box is typically limited to 10-20
particles. Optical or confocal microscopy of colloidal suspensions
or imaging of 2-dimensional driven granular materials could in
principle address this issue. In practice, however, the number of
particles that can be followed is limited. This is because the
spatial resolution is a fixed fraction of the field of view in
conventional imaging techniques.  Near jamming, however, the
particle displacements are very restrained, such that a high
magnification has to be chosen to resolve them; this consequently
limits the field of view and thus the number of probed particles.

Here, we overcome these limitations by introducing a novel optical
method that combines features of both dynamic light
scattering~\cite{Berne} and imaging. The method, termed Photon
Correlation Imaging (PCIm)~\cite{DuriPhD}, decouples the length
scale over which the dynamics is probed from the size of the field
of view, thereby allowing long-ranged correlations to be measured.
We demonstrate PCIm by measuring the range, $\xi$, of spatial
correlations in the dynamics of three systems: a diluted
suspension of Brownian particles, and two jammed materials, a dry
foam and a colloidal gel. As expected, we find no spatial
correlations in the dynamics of the Brownian particles. For the
foam consisting of deformable bubbles that can easily slip around
each other, $\xi$ is of the order of a few bubble sizes, but
partially correlated motion is also observed on larger length
scales. For the gel consisting of rigidly bound hard colloids, the
range of correlation is comparable to the system size, thousands
of times larger than the particle size.

The Brownian particles are polystyrene spheres of radius $265$ nm,
suspended in a mixture of water and glycerol at a volume fraction
$\varphi \sim 10^{-4}$. The foam is a commercial shaving foam
consisting of tightly packed, polydisperse bubbles (gas fraction
$\approx 0.92$), a model system used to study intermittent
dynamics in previous investigations~\cite{DurianScience1991}. To
avoid multiple scattering, we confine the foam in a thin cell, of
thickness 1-2 bubble layers. The foam coarsens with time: the data
reported here are taken 6.1 h after producing the foam, when the
average bubble radius is $\sim80~\mu$m. As an example of a system
where jamming results from attractive interactions, we study a gel
made of aggregated polystyrene particles of radius 20 nm at
$\varphi = 6 \times 10^{-4}$, 32.3 h after the gel was formed.
Details of the system can be found
in~\cite{LucaPRL2000,DuriEPL2006}; its average dynamics are
representative of those of a wide class of soft materials. We
express all time scales in units of $\taur$, the relaxation time
of the average dynamics, defined by $f(\taur) = \mathrm{e}^{-1}$,
with $f(\tau)$ the usual average dynamic structure factor. For the
Brownian particles, foam, and gel, $\taur = 42, 140,$ and 5000 s,
respectively. For all samples, the dynamics are stationary on the
time scale of the experiments.

\begin{figure}
\epsfig{file=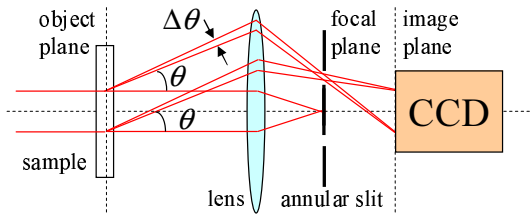,width=7cm} \caption{(Color online)
Schematic view of the low-angle, space-resolved light scattering
apparatus.} \label{fig:apparatus}
\end{figure}

A scheme of the space-resolved dynamic light scattering apparatus
is shown in Fig.~\ref{fig:apparatus}. The sample cell of thickness
$L = 2$ or 0.1 mm for colloids and foams, respectively, is placed
in a temperature-controlled bath (not shown) and is illuminated by
a collimated laser beam with a diameter of~8 mm and a wavelength
$\lambda = 633$ nm. A lens with a focal length $ f_\mathrm{L} =
40$ mm is used to form an image of the sample onto the detector of
a CCD camera, typically with magnification 0.8-1.3. We place an
annular aperture of radius $r_\mathrm{a} = 9$ mm in the focal
plane of the lens, so that only light scattered within a small
angular interval $\Delta \theta /\theta \approx 0.1$ around
$\theta \approx 6.4$ deg contributes to the formation of the
image. The images have a speckled appearance; each speckle
receives light scattered from a small volume of lateral extension
$\approx \lambda/\Delta \theta$~\cite{Goodman2007} and depth $L$.
Our setup combines features of both imaging and scattering
methods: similarly to imaging techniques and in contrast to usual
scattering methods, each location on the CCD detector corresponds
to a precise location in the sample. However, similarly to
scattering methods and unlike conventional imaging, only light
scattered at a chosen angle and thus scattering vector $q$ is
collected by the detector, allowing one to probe the dynamics of
the sample on a well defined length scale $\Lambda \approx q^{-1}
\approx 1~\mu\mathrm{m}$, optimized according to the sample
features.

Speckle correlography~\cite{Goodman2007} and setups similar to the
one in
Fig.~\ref{fig:apparatus}~\cite{Williams1996,SchopeLangmuir2006}
have been used in the past to study structural properties. Here we
focus on dynamical properties. We take a time series of CCD images
at a fixed rate; each image is divided in regions of interest
(ROIs), typically squares of side $10-20$ pixels containing $\sim
100$ speckles and corresponding to about
$50^2-150^2~\mu\mathrm{m}^2$ in the sample. The local dynamics
within a given ROI is quantified by a two-time degree of
correlation~\cite{LucaJPCM2003}:

\begin{eqnarray}
c_I(t,\tau;\mathbf{r}) = \frac{\left < I_p(t)I_p(t+\tau) \right
>_{\mathrm{ROI}(\mathbf{r})} }{\left < I_p(t) \right
>_{\mathrm{ROI}(\mathbf{r})} \left < I_p(t+\tau) \right
>_{\mathrm{ROI}(\mathbf{r})} } -1 \,,
\label{eq:cI}
\end{eqnarray}
with $\mathbf{r}$ the position of the center of the ROI, $I_p(t)$
the intensity of the $p$-th pixel of the ROI at time $t$, and $\left
< \cdot \cdot \cdot \right
>_{\mathrm{ROI}(\mathbf{r})}$ the average taken over all pixels
within the ROI. The space and time resolved dynamic structure
factor is given by $f(t,\tau;\mathbf{r}) = \sqrt{\beta^{-1}
c_I(t,\tau;\mathbf{r})}$, with $\beta \lesssim 1$ an instrumental
constant~\cite{Berne} and $f(t,\tau;\mathbf{r}) =
N^{-1}\sum_{j,k}\exp[i\mathbf{q} \cdot
(\mathbf{r}_j(t)-\mathbf{r}_k(t+\tau))]$, where the sum is over
the $N$ scatterers belonging to the sample volume associated to
the ROI. The usual intensity correlation function $g_2(\tau)-1$
measured in traditional light scattering is the average of
$c_I(t,\tau;\mathbf{r})$ over both $t$ and $\mathbf{r}$. By
applying Eq.~(\ref{eq:cI}) to all ROIs, we build a ``dynamical
activity map'' (DAM) showing the local degree of correlation,
characterizing the change in configuration during the time
interval $[t,t+\tau]$.

\begin{figure}
\epsfig{file=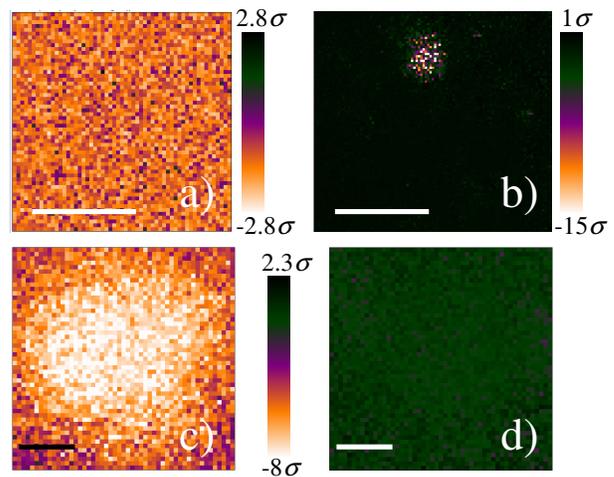,width=8cm} \caption{(Color online) Dynamic
activity maps; the bars correspond to 2 mm. The color
code indicates, in units of standard deviations $\sigma$, the
fluctuations of the local $c_I$ with respect to its temporal
average. a): Brownian particles ($\tau = 0.2$ s, $\tau/\taur =
0.0048$); b) foam ($\tau= 3$ s, $\tau/\taur =
0.021$); c) and d): colloidal gel ($\tau = 500$ s, $\tau/\taur =
0.1$) during a rearrangement event and a quiescent period,
respectively.} \label{fig:dam}
\end{figure}

In Fig.~\ref{fig:dam} we show representative DAMs for the three
samples; movies displaying sequences of DAMs can be found in
\cite{SOM}. While previous works on dynamical heterogeneity have
often dealt with the spatial correlation of the dynamics on a time
scale comparable to the average relaxation time
$\tau_r$~\cite{DauchotPRL2005_2,Durian,LacevicPRE}, here we focus
on smaller time lags, for which individual rearrangement events
are best visualized. As a control test, we show in
Fig.~\ref{fig:dam}a a DAM for a suspension of Brownian particles.
The dynamics exhibit spatial fluctuations, without any obvious
spatial correlation, consistent with the behavior expected for
diluted, non-interacting scatterers. A typical DAM for the foam
exhibiting an intermittent rearrangement event is shown in
Fig.~\ref{fig:dam}b. Contrary to the case of the Brownian
particles, the dynamics is inhomogeneous: the bright spot at the
top of the field of view corresponds to an extended region with a
large loss of correlation, while the rest of the sample
experiences only a marginal change in configuration. The DAM
movie~\cite{SOM} reveals that the foam dynamics is dominated by
similar events. We identify these intermittent, localized events
with the bubble rearrangements driven by the accumulation of
internal stress due to coarsening. Spatial maps of the dynamics of
the colloidal gel are shown in Fig.~\ref{fig:dam}c and d.
Similarly to the foam, the dynamics is due to intermittent
rearrangement events (c), separated by quiescent periods (d).
These events, however, extend over strikingly large length scales,
comparable to the system size, suggesting that any local
rearrangement must propagate very far throughout the network.

\begin{figure}
\epsfig{file=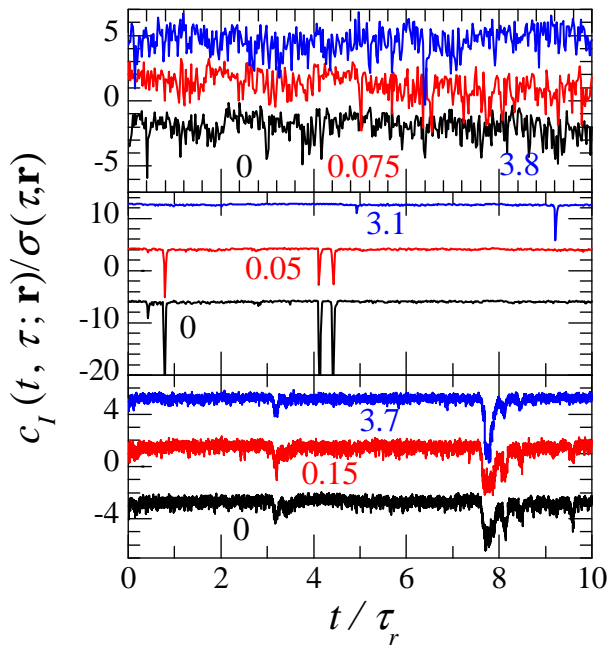,width=8cm} \caption{(Color online)
Instantaneous degree of correlation $c_I(t,\tau,\mathbf{r})$,
normalized by its standard deviation $\sigma(\tau,\mathbf{r})$,
measured at three different locations in the sample, as a function
of reduced time $t/\taur$ (for the sake of clarity, the curves are
offset along the vertical axis and only a limited portion of the
data is shown). Data have been corrected for instrumental noise as
explained in Ref.~\cite{DuriPRE2005}. For each sample, curves are
labeled by the distance in mm with respect to the location for the
black curve. Top: Brownian particles (the labels 0, 0.075 and 3.8
refer to the traces from bottom to top) ; middle: foam; bottom:
colloidal gel. For all samples, $\tau/\taur$ is as in
Fig.~\ref{fig:dam}.} \label{fig:cI}
\end{figure}

To better evaluate the spatio-temporal correlations of the
dynamics, we compare in Fig.~\ref{fig:cI} the time evolution of
the instantaneous degree of correlation $c_I(t,\tau;\mathbf{r})$
in three ROIs, two of which are adjacent, while the third one is
at least 3.1 mm apart from the first one. Data for the Brownian
particles are shown in the top panel. The local dynamics
fluctuates in a noise-like way and fluctuations in distinct ROIs
appear to be uncorrelated, whatever the distance between ROIs. For
the foam (middle panel), individual rearrangement events such as
the one observed in Fig.~\ref{fig:dam}b result in deep downward
spikes, lasting $\sim3$ s, departing from the average value of
$c_I$ by several standard deviations. The depth of the spikes is a
measure of the displacement of  the scatterers compared to
$\Lambda$. The $c_I$ traces for adjacent ROIs appear to be
strongly correlated, while the spikes in the trace of the third
ROI are almost completely uncorrelated, consistently with the
typical event size inferred from the DAM of Fig.~\ref{fig:dam}b.
For the colloidal gel (bottom panel), we find that essentially all
ROIs present strongly correlated $c_I$ traces, suggesting that the
ultra-long ranged rearrangement events depicted in
Fig.~\ref{fig:dam}c are indeed representative of the typical
behavior of the gel. Their duration is of the order of $10^3$ s.

\begin{figure}
\epsfig{file=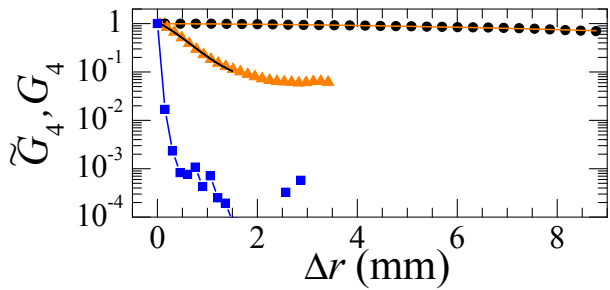,width=8cm} \caption{(Color
online) Spatial correlation of the dynamics for diluted Brownian
particles ($\widetilde{G_4}$, blue squares), for a foam ($G_4$,
orange triangles), and for a colloidal gel ($G_4$, black circles).
The lines are stretched exponential fits to the decay of $G_4$,
yielding $\xi = 0.6$ and 16.4 mm for the foam and the gel,
respectively. For all samples, $\tau/\taur$ is as in
Fig.~\ref{fig:dam}.} \label{fig:spcorrelation}
\end{figure}

We quantify the spatial correlation of the dynamics by introducing a
``four point'' correlation function $\widetilde{G}_4(\Delta r,\tau)$
that compares the dynamical activity in two small regions separated
by $\Delta r$. We define $\widetilde{G}_4$ as the crosscorrelation
of the local
dynamics~\cite{GlotzerReview,DauchotPRL2005_2,Durian,LacevicPRE}:
\begin{eqnarray}
\widetilde{G}_4(\Delta r,\tau) = \left < \frac{ \left < \delta
c_I(t,\tau;\mathbf{r}_1)\delta c_I(t,\tau;\mathbf{r}_2) \right
>_t }{\sigma(\tau,\mathbf{r}_1) \sigma(\tau,\mathbf{r}_2)} \right
>_{|\mathbf{r}_1-\mathbf{r}_2|=\Delta r}.
\label{eq:G4}
\end{eqnarray}
Here, $\delta c_I(t,\tau;\mathbf{r}) = c_I(t,\tau;\mathbf{r}) -
\left< c_I(t,\tau;\mathbf{r}) \right >_t$ and
$\sigma(\tau,\mathbf{r}) = \sqrt{\left< \delta
c_I(t,\tau;\mathbf{r})^2 \right >_t}$, with $\left< \cdot \cdot
\cdot \right>_t$ an average over time.
Figure~\ref{fig:spcorrelation} shows $\widetilde{G}_4$ for the
Brownian particles (blue squares). The spatial correlation decays
nearly completely over one ROI size, confirming that the dynamics of
diluted Brownian particles are spatially uncorrelated. The spatial
correlation of the dynamics of the foam and the gel are shown in
Fig.~\ref{fig:spcorrelation} as orange triangles and black circles,
respectively. Here, a slightly different normalization has been
chosen: as discussed in Ref.~\cite{DuriPRE2005}, $c_I$ contains a
noise contribution stemming from the finite number of speckles. The
noise of distinct ROIs is uncorrelated and thus does not contribute
to the numerator of Eq.~(\ref{eq:G4}); however, it does add an extra
contribution for $\Delta r = 0$, and it contributes to the $\sigma$
terms in the denominator. We thus plot only data for $\Delta r> 0$
and define $G_4(\Delta r,\tau) = b\widetilde{G}_4(\Delta r,\tau)$,
with $b \gtrsim 1$ chosen so that $G_4(\Delta r,\tau) \rightarrow 1$
for $\Delta r \rightarrow 0$.

For the foam, most of the decay of $G_4$ is well described by a
slightly compressed exponential, $G_4 = \exp[(-\Delta R/\xi)^p]$,
with $p=1.27$ and $\xi = 0.6~\mathrm{mm}$ (black line),
corresponding to 8 bubble radii, denoting the size of rearrangement
events, in agreement with previous
estimates~\cite{DurianScience1991}. Interestingly, this decay is
followed by a plateau $G_4 \approx 7\times10^{-2}$ that extends over
the full field of view, $\Delta r \lesssim 4$ mm. This indicates
that, although the spatial correlations of the dynamics are
dominated by the typical size of rearrangements, the events have
also a small impact on the rest of the sample. A comparison with
younger foams, where the confinement is less severe, suggests that
this long-range strain propagation is due to the strong confinement
of the sample investigated. Elastic strain propagation has a
dramatic effect in the gel, where the range of dynamical correlation
is strikingly large: $G_4$ only decays by about 25\% over 8 mm. A
compressed exponential fit of $G_4$ (red line) yields $p= 1.8$ and
$\xi = 16.4~\mathrm{mm}$, demonstrating that the dynamics are
strongly correlated over distances far larger than any relevant
structural length scale: the particle diameter is 20 nm and the gel
is formed by fractal clusters of size $\approx
20~\mu\mathrm{m}$~\cite{LucaPRL2000}. The differences observed in
the spatial correlations of our two jammed systems can be understood
considering the differences in the response to a local change of
configuration. The foam consists of deformable bubbles, for which
the resistance to sliding around each other is relatively small. Any
local rearrangement is then mostly dissipative, involving the
topological rearrangement of a few bubbles, and setting only a small
elastic strain on the surrounding medium. By contrast, in the
colloidal gel the rigid connection of particles to a stress-bearing
backbone entails a long-ranged strain field, as both the particles
and the bonds between them are rigid and do not allow for a
localized reorganization.

Our experiments show that the dynamics of jammed systems can
exhibit surprisingly long-ranged spatial correlations, which
exceed by far those reported in previous works on molecular and
colloidal glass formers and in driven granular
media~\cite{EdigerReview,RichertReview,GlotzerReview,WeeksScience2000,DauchotPRL2005_2,BerthierScience2005,PREAlba,Durian,LacevicPRE}.
The technique introduced here should provide a flexible and
valuable tool to characterize spatial correlations of the
dynamics. PCIm-experiments can be performed in any of the
classical light scattering geometries, including small angle
(where $\Lambda$ can be tuned by varying
$r_{\textrm{a}}/f_{\textrm{L}}$), wide angle (see Fig. SM1
in~\cite{SOM}), and diffusing wave spectroscopy in the
backscattering geometry~\cite{dws}, thereby providing access to a
range of length-scales spanning more than 1.5 decades. Ongoing
wide angle PCIm experiments performed on
Laponite~\cite{someLaponiteref}, closely packed
onions~\cite{RamosPRL2001}, and a fibrin film~\cite{FibrinRef}
suggest that the long range dynamical correlations observed here
for the gels may be very general. In the future, it would be
interesting to compare systematically dynamical correlations in
jammed (connected) systems and glassy (disconnected) systems.
Finally, we stress that PCIm could also be extended to different
kinds of radiation, to probe for instance molecular glass formers.
Indeed, the optical layout shown Fig.~\ref{fig:apparatus} is
similar to the one used in Fluctuation Electron Microscopy
(FEM)~\cite{FEM} and the PCIm method could be adopted to analyze
the FEM speckle images to monitor the dynamics with spatial
resolution.

This work was supported by CNES, ACI JC2076, CNRS (PICS n. 2410),
and the Swiss National Science Foundation. L.C. acknowledges
support from the Institut Universitaire de France.


\end{document}